\documentclass[12pt,a4paper]{article}
\pdfoutput=1

\usepackage[utf8]{inputenc}

\usepackage{multirow}
\usepackage{cite}
\usepackage{amsmath}
\usepackage{color}
\usepackage{amsfonts}
\usepackage{amssymb}
\usepackage{graphicx}
\usepackage{geometry}
\usepackage{amssymb,epsfig,subfigure}
\usepackage{hyperref}
\usepackage{url}
\usepackage{comment}
\usepackage[font=footnotesize]{caption}
\usepackage{slashed}
\usepackage{bm}

\makeatletter
\renewcommand\section{\@startsection {section}{1}{\z@}%
                                 {-3.5ex \@plus -1ex \@minus -.2ex}
                                   {2.3ex \@plus.2ex}%
                                   {\normalfont\large\bfseries}}
\renewcommand\subsection{\@startsection{subsection}{2}{\z@}%
                                   {-3.25ex\@plus -1ex \@minus -.2ex}%
                                     {1.5ex \@plus .2ex}%
                                     {\normalfont\bfseries}}
\renewcommand\subsubsection{\@startsection{subsubsection}{3}{\z@}%
                                   {-3.25ex\@plus -1ex \@minus -.2ex}%
                                     {1.5ex \@plus .2ex}%
                                     {\normalfont\itshape}}
\makeatother

\def\pplogo{\vbox{\kern-\headheight\kern -29pt
\halign{##&##\hfil\cr&{\ppnumber}\cr\rule{0pt}{2.5ex}&\ppdate\cr}}}
\makeatletter
\def\ps@firstpage{\ps@empty \def\@oddhead{\hss\pplogo}%
  \let\@evenhead\@oddhead 
}
\def\maketitle{\par
 \begingroup
 \def\thefootnote{\fnsymbol{footnote}}
 \def\@makefnmark{\hbox{$^{\@thefnmark}$\hss}}
 \if@twocolumn
 \twocolumn[\@maketitle]
 \else \newpage
 \global\@topnum\z@ \@maketitle \fi\thispagestyle{firstpage}\@thanks
 \endgroup
 \setcounter{footnote}{0}
 \let\maketitle\relax
 \let\@maketitle\relax
 \gdef\@thanks{}\gdef\@author{}\gdef\@title{}\let\thanks\relax}
\makeatother

\numberwithin{equation}{section}

\newcommand\eea{\end{eqnarray}}
\newcommand\bea{\begin{eqnarray}}

\def\beq{\begin{equation}}
\def\eeq{\end{equation}}
\def\d{\partial}

\def\d{\partial}

\newcommand{\be}{\begin{equation}}
\newcommand{\ee}{\end{equation}}
\newcommand{\ba}{\begin{align}}
\newcommand{\ea}{\end{align}}
\newcommand{\bg}{\begin{gather}}
\newcommand{\eg}{\end{gather}}
\newcommand{\bseq}{\begin{subequations}}
\newcommand{\eseq}{\end{subequations}}

\newcommand{\tr}{{\rm tr}}
\newcommand{\mc}{\mathcal}

\newcommand{\dd}[1]{\text{d}#1}

\newcommand{\of}[1]{\left(#1\right)}
\newcommand{\off}[1]{\left[#1\right]}

\newcommand{\abs}[1]{\left| #1 \right|}

\newcommand{\coment}[1]{}

\begin{document}
\setcounter{page}0
\def\ppnumber{\vbox{\baselineskip14pt
}}
\def\ppdate{
} \date{}

\author{Nicolás Abate$^{1, 2 }$, Horacio Casini$^{1, 2 }$, Marina Huerta$^{1, 2 }$, Leandro Martinek$^{1, 2}$\\
[7mm] \\
{\normalsize \it $^1$Centro At\'omico Bariloche and CONICET}\\
{\normalsize \it $^2$ Instituto Balseiro, UNCuyo and CNEA}\\
{\normalsize \it S.C. de Bariloche, R\'io Negro, R8402AGP, Argentina}\\
}

\bigskip
\title{\bf  
Exact Mutual Information Difference: 

Scalar vs. Maxwell Fields
\vskip 0.5cm}
\maketitle

\begin{abstract}

We compute, for any Rényi index 
$n$, the exact difference between the mutual Rényi informations of a pair of free massless scalars and that of a Maxwell field in 
$d=4$ dimensions. Using the standard dimensional reduction method in polar coordinates, the problem is mapped to that of a single scalar field in 
$d=2$ with Dirichlet boundary conditions, which in turn can be conveniently related to the algebra of a chiral current on the full line. This latter identification, which maps algebras on an interval to two-interval algebras, yields exact results that clarify the structure of the long-distance OPE perturbative expansion of the mutual information. We find that this series has a finite radius of convergence only for integer 
$n>1$, while it becomes only asymptotical for 
$n=1$ and general non-integer values of $n$.

\end{abstract}
\bigskip

\newpage

\tableofcontents


\section{Introduction}

The mutual information $I(A,B)$ between two disjoint regions $A,B$, is an interesting information measure in QFT. It quantifies both classical and quantum correlations, and when the distance between the boundaries of $A$ and $B$ is small, it serves as a universal geometric regularization of the entanglement entropy \cite{Casini:2015woa}. Thus, at short distances it contains information about the renormalization group charges. In particular, the coefficient of the Euler term in the trace anomaly appears as the coefficient of the logarithmic term in even dimensions. On the other hand, at large distances, the mutual information has an expansion computable in terms of the boostrap data of a conformal field theory (CFT) \cite{Calabrese:2010he,Cardy:2013nua}. This suggests the control of this series expansion may have a clue to understand Cardy-type formulas relating central charges to boostrap data in higher dimensions.    

The replica trick and the method of operator product expansion (OPE) of replica twists  start with Rényi entropies of integer index $n$.  Given some region $A$, its $n$-th Rényi entropy is given by
\be\label{eq:Renyi_def}
S_n(A)=(1-n)^{-1}\log\tr\rho_A^n\,,
\ee
where $\rho_A$ is the density matrix of the vacuum state reduced to the region $A$. Then, the $n$-th Rényi mutual information (RMI) between a pair of regions $A$ and $B$ is defined as the regulator free combination
\be\label{eq:RMI_def}
I_n(A,B)=S_n(A)+S_n(B)-S_n(A\cup B)\,.
\ee
In the limit $n\to1$, since the expression \eqref{eq:Renyi_def} reduces to the entanglement entropy (EE) for the region $S(A)$, the formula \eqref{eq:RMI_def} yields the mutual information (MI)
\be
\lim_{n\to1}I_n(A,B)\equiv I(A,B)=S(A)+S(B)-S(A\cup B)\,.
\ee
For a CFT, it is natural to consider spherical regions. We will take these spheres lying on the plane $x^0=0$ in Minkowski space. In this case, the RMI are functions of a single cross ratio, $I_{n}(A,B) = I_n(\eta)$, where 
\be
\eta= \frac{R_A\, R_B}{(R_A + L)(R_B + L)}\in (0,1)\,,\label{eta}
\ee
and $R_A,R_B$ are the radius of the two spheres, while $L$ the separation distance. 

Substantial effort has been made to compute the function $I_n(\eta)$ in different models. The known exact results corresponds to $1+1$ dimensions where the replica twists can be written in terms of field operators in the replica orbifold. Rényi entropies for integer $n$ where computed for several $1+1$ CFT's \cite{Calabrese:2009ez,Calabrese:2010he} and BCFT's \cite{Mintchev:2020uom, Estienne:2023ekf}. The case of $n=2$ is specially treatable since the replica partition function can be mapped to the torus partition function \cite{furukawa2009mutual,Headrick:2010zt}. However, in most cases it is not known how to make the analytic continuation to non integer $n$, in particular to the case of the mutual information $n=1$. Exact results for the mutual information $I(\eta)$ for two intervals in $1+1$ dimensions are only known for free fields \cite{casini2005entanglement,Arias:2018tmw}. In this case the functions $I_n(\eta)$ are also known for any $n$. 

In higher dimensions no exact result is known for any model and any $n$. For holographic theories the leading large $N$ term can be computed by the Ryu-Takayanagi formula for $n=1$ \cite{hirata2007ads}. This, however, has a phase transition in $\eta$ and vanishes to all orders in the large distance limit. Substantial progress has been made in understanding the large distance expansion in the general case \cite{Calabrese:2010he,Cardy:2013nua,agon2016quantum,agon2025mutual,Chen:2016mya,chen2013short,chen2017mutual,Casini:2021raa,agon2022tripartite}. 

In this paper we compute exactly the difference between (twice) the RMI for a scalar field $I_n^{(\text{S})}(\eta)$ and the one for a Maxwell field $I_n^{(\text{M})}(\eta)$, for any $n$, in $d=4$. The mutual information for the Maxwell field has been studied previously in relation with the problem of the anomaly mismatch (see \cite{casini2020logarithmic} and references therein). The free Maxwell field has a logarithmic coefficient that does not match the expected result proportional to the conformal anomaly. It was found that this mismatch is due to the non local sectors of the Maxwell field (conservation of electric and magnetic flux) and is repaired if the theory of the Maxwell field is completed by coupling it with heavy electric and magnetic charges \cite{casini2020logarithmic}.    

Decomposing the free Maxwell field in polar coordinates it was found that the operator algebras for spherical regions coincide with the ones of a pair of free massless scalars, except for the mode of zero angular momentum $\ell=0$ that is missing in the Maxwell field \cite{Casini:2015dsg}.  This mode is found to be equivalent to the theory of a free massless scalar on the half-line in $d=2$ with Dirichlet boundary conditions at the origin. In this paper we further map the algebras of the latter theory to the ones of a chiral current in $d=2$ depending on the type of region considered, as follows:
\be
\begin{aligned}
    \mc A_\mathrm{HL}\big((a,b)\big) &\longrightarrow\, \mc A_j^{\max}\big((-b,-a)\cup(a,b)\big)\\
    \mc A_\mathrm{HL}\big((0,a)\cup(b,\infty)) &\longrightarrow\, \mc A_j^{\text{add}}\big((-\infty,-b)\cup(-a,a)\cup(b,\infty)\big)\,.
\end{aligned}
\ee
Here $0<a<b$; $\mc A_{\text{HL}}$ and $\mc A_j$ are the algebras for the scalar on the half-line and the chiral current, respectively; and the superscripts `max' and `add' refer to the maximal or additive choice for these algebras \cite{Casini:2019kex,Casini:2021zgr}, as explained below. With this identification at hand, we obtain an identity between RMI for theories of different dimensionality:
\be
2 I_n^{(\text{S})}(\eta)-I_n^{(\text{M})}(\eta)=2 I_n^{(\text{C})}(\eta)\,,
\ee
where $I_n^{(\text{C})}(\eta)$ is the RMI for the chiral current field in $d=2$. This last two-dimensional function was computed in \cite{Arias:2018tmw}. Therefore, we obtain an exact result for the mutual information {\sl difference} between two four dimensional models. We check several known results for the long distance expansion of this function, and obtain some new coefficient for the contribution of a replica operator for the Maxwell field. Having the complete expression, we verify wether the long distance expansion series is convergent, finding an intriguing result: the series is convergent for integers $n>1$ while it is only asymptotic for other values of $n$.   

This article is organized as follows. In section \ref{maxwell} we review the polar decomposition of the Maxwell field and its relation with the scalar field in $d=4$ dimensions. In section \ref{sec:2} we introduce the scalar field on the half-line, as well as the chiral current on the whole null-line. These theories turn out to be equivalent and in this section we also establish this relation in terms of operator algebras. This result is essential in section \ref{sec:3} to find the difference of RMIs between scalar and Maxwell fields, or equivalently the RMI for the scalar field on the half-line, since the RMI for the chiral current is already known as a function of the cross-ratio  \cite{Arias:2018tmw}. We support our results with numerical calculations on a lattice and also discuss some limits when the intervals involved are very near or far apart from each other. Then, we examine the convergence of the long distance OPE expansion for the RMI in section \ref{sec:33}. We conclude with a brief summary of our results and comment on some potentially interesting future lines of study in section \ref{sec:4}.

\section{Maxwell versus scalar field in $d=4$}
\label{maxwell}

By exploiting the spherical symmetry, the Maxwell theory in $d=4$ dimensions can be decomposed into independent one-dimensional radial sectors labeled by the angular momentum quantum numbers $(\ell,m)$. This dimensional reduction establishes a direct correspondence with the massless scalar theory in four dimensions for spherical geometries. This was done in detail in \cite{Casini:2015dsg} to compare the respective logarithmic coefficients in the EE. Here we briefly review this relation. For this purpose, we expand the electric and magnetic fields in vector spherical harmonics $\mathbf{Y}_{\ell m}^{s}$ with $s=r,e,m$, that stand for the ``radial'', ``electric'' and ``magnetic'' type of vector harmonics, respectively. This allows us to analyze each radial sector independently:
\be
\begin{aligned}
\mathbf{E} &= \sum_{\ell,m} \of{E_{\ell m}^{r}(r) \mathbf{Y}_{\ell m}^{r} + E_{\ell m}^{e}(r) \mathbf{Y}_{\ell m}^{e} + E_{\ell m}^{m}(r) \mathbf{Y}_{\ell m}^{m}}\,,\\
\mathbf{B} &= \sum_{\ell,m} \of{B_{\ell m}^{r}(r) \mathbf{Y}_{\ell m}^{r} + B_{\ell m}^{e}(r) \mathbf{Y}_{\ell m}^{e} + B_{\ell m}^{m}(r) \mathbf{Y}_{\ell m}^{m}}\,.
\label{eq:EBlmexp}
\end{aligned}
\ee
The vector harmonics are related to the standard scalar spherical harmonics $Y_{\ell m}$. For the sector $\ell=0$, only the radial harmonic $\mathbf{Y}_{00}^{(r)}=Y_{00}\,\hat r$ is defined. For the sectors with $\ell\geq1$, the relation is given by
\begin{equation}
\mathbf{Y}_{\ell m}^{r} =Y_{\ell m}\, \hat{r}\,, \quad 
\mathbf{Y}_{\ell m}^{e}=\frac{r\,\bm{\nabla} Y_{\ell m}}{\sqrt{\ell(\ell+1)}}\,,\quad 
\mathbf{Y}_{\ell m}^{m} = \frac{\mathbf{r} \times \bm{\nabla} Y_{\ell m}}{\sqrt{\ell(\ell+1)}}\,,
\end{equation}
with the normalization and orthogonality condition
\begin{equation}
\int \dd\Omega \ \mathbf{Y}_{\ell m}^{s}(\theta,\varphi) \cdot \mathbf{Y}_{\ell' m'}^{s'}(\theta,\varphi) = \delta_{s s'} \, \delta_{\ell \ell'} \, \delta_{m m'}\,, \qquad s,s' = r,e,m\,.
\label{eq:ortho}
\end{equation}

Using the expansion \eqref{eq:EBlmexp} the Hamiltonian decomposes as a sum of independent $(\ell,m)$ sectors. This is $H=\sum_{\ell\geq1,m}H_{\ell m}$, where 
\begin{equation}
H_{\ell m} = \frac{1}{2} \int \dd r \ r^2 \sum_{s=r,e,m} \Big[ (E_{\ell m}^{s})^2 + (B_{\ell m}^{s})^2 \Big]\,, \qquad \ell\geq1\,.
\label{eq:Hlm}
\end{equation}
The constraints $\nabla\cdot\mathbf{E}=\nabla\cdot\mathbf{B}=0$ force the contribution of the mode $\ell=0$ to vanish and  relate the electric components  $E_{\ell m}^{e}$ and $B_{\ell m}^{e}$ to the radial ones. Then, after a suitable rescaling that render the canonical commutation relations standard, one arrives at two sets of modes $(\tilde E_{\ell m}^{r},\tilde B_{\ell m}^{m})$ and $(\tilde B_{\ell m}^{r},\tilde E_{\ell m}^{m})$ identical and independent, where $\tilde{E}_{\ell m}^{r},\tilde{B}_{\ell m}^{r}$ and $\tilde{E}_{\ell m}^{m},\tilde{B}_{\ell m}^{m}$ represent the rescaled canonical variables corresponding to the radial and magnetic components, respectively. The Hamiltonian for each sector then takes the form
\begin{equation}
H_{\ell m} = \frac{1}{2}\int_0^\infty \dd r\  \Big[ (\partial_r \tilde{E}_{\ell m}^{r})^2 + \frac{\ell(\ell+1)}{r^2}(\tilde{E}_{\ell m}^{r}) ^2 + (\tilde{B}_{\ell m}^{m})^2 \Big]+(\tilde E_{\ell m}\leftrightarrow \tilde{B}_{\ell m}) \,,
\end{equation}
with
\be
[\tilde{E}_{\ell m}^r(r), \tilde{B}^m_{\ell'm'}(r')]=i\,\delta_{\ell,\ell'}\delta_{m,m'}\delta(r-r')\,,\quad [\tilde{B}_{\ell m}^r(r), \tilde{E}^m_{\ell'm'}(r')]=i\,\delta_{\ell,\ell'}\delta_{m,m'}\delta(r-r')\,.
\ee

Note that each term in the last expression has precisely the same Hamiltonian structure as that of a massless scalar field in four dimensions decomposed into spherical modes. For comparison, consider a massless scalar field $\phi=\sum_{\ell\ge 0, m}\phi_{\ell m}(r)Y_{\ell m}(\theta,\varphi)$. By introducing $\tilde{\phi}_{\ell m}=r\phi_{\ell m}$, the radial Hamiltonian reads
\begin{equation}
H^{(\text{S})}_{\ell m} = \frac{1}{2}\int_0^\infty \dd r\ \Big[ \tilde{\pi}_{\ell m}^2 + (\partial_r \tilde{\phi}_{\ell m})^2 + \frac{\ell(\ell+1)}{r^2}\tilde{\phi}_{\ell m}^2 \Big]\,, \label{pot}
\ee 
with 
\be
\quad [\tilde{\phi}_{\ell m}(r),\tilde{\pi}_{\ell'm'}(r')]=i\, \delta_{\ell,\ell'}\delta_{m,m'}\delta(r-r')\,.
\end{equation}
This is identical to each copy of the electromagnetic Hamiltonian above, except for the presence of the additional $\ell=0$ scalar mode. Hence, the Maxwell field in spherical geometry is equivalent to two copies of a massless scalar field with the $\ell=0$ mode removed. Interestingly, the missing mode does not have the potential term in (\ref{pot}). Therefore it behaves as a massless scalar in $d=2$ defined on the half-line with Dirichlet boundary conditions at the origin. This is special for the dimensional reduction of the scalar field in $d=4$, because the mass term for other dimensions does not coincide with that of the Maxwell field (see Eq. (25) in \cite{Huerta:2022tpq}).

Since the total entropy is an additive sum of the independent contributions of each angular momentum sector, the entanglement entropy of a free Maxwell field on any number of spherical shells is equivalent to that of two massless scalar fields with the $\ell=0$ mode removed in the same geometry. The geometry of two disjoint balls can be conformally mapped to the one of a ball $A$ of radius $a$ and the complement $B$ of a bigger concentric ball of radius $b>a$. Then, the cross ratio (\ref{eta}) in terms of $a,b$ reads
\be\label{eta0}
\eta=\frac{4ab}{(a+b)^2}\,.
\ee

Then, the above identification for spherical geometries implies the following relation between the RMIs of these three theories:
\be\label{eq:diff_RMI}
2 I_n^{(\text{S})}(\eta)-I_n^{(\text{M})}(\eta)=2 I_n^{(0)}(\eta)\,,
\ee
where we have defined the following quantity for the $d=2$ scalar with Dirichlet boundary condition (the $\ell=0$ mode of the $d=4$ scalar)
\be
I^{(0)}_n(\eta)=S_n((0,a))+S_n((b,\infty))-S_n((0,a)\cup (b,\infty))\,.
\ee

\section{Scalar with Dirichlet boundary condition in $d=2$}\label{sec:2}

We begin this section by introducing the theory of a massless scalar field on the half-line in $d=2$ satisfying Dirichlet boundary conditions at the origin. This model will turn out to be equivalent to a chiral scalar field living on the null-line, so this theory is also introduced. The relation between both is later established in terms of operator algebras.

\subsection{The scalar field and the chiral current}

The action for the scalar on the half-line is given by
\be\label{eq:S_dirichlet}
S=\frac{1}{2}\int_{\mathbb{R}}\dd t\int_0^\infty\dd x\off{-\of{\partial_t\phi}^2+\of{\partial_x\phi}^2},
\ee
for $\phi$ satisfying Dirichlet boundary conditions $\phi(0)=0$. Quantizing the field at the $t=0$ slice and imposing canonical commutation relations between $\phi$ and its conjugate momentum $\pi$, one arrives at the following form for the two-point functions
{\small
\be
    \langle\phi(x)\phi(y)\rangle=\log\left|\frac{x+y}{x-y}\right|,
    \quad
    \langle\phi(x)\pi(y)\rangle=\frac{i}{2} \delta(x-y)
    \quad   
    \text{and}  \quad   
    \langle\pi(x)\pi(y)\rangle=-\frac{2}{\pi}\frac{xy}{\of{x^2-y^2}^2}\,.
\label{eq:halflinescalardef}
\ee}
By gaussianity, these functions carry the whole information of the theory.

On the other hand, the theory of a chiral current $j$ can be obtained from a massless field $\varphi$ in $d=2$ via its null derivative, this is $j(x_+)=\partial_{x_+}\varphi(t,x)$ where $x_+=x+t$. Then, this model is defined by the two-point function
\be\label{eq:j_corr}
\langle j(x_+)j(y_+)\rangle=-\frac{1}{2\pi}\frac{1}{\of{x_+-y_+-i\epsilon}^2}\,,
\ee
and the commutation relation
\be\label{eq:j_comm}
[j(x_+),j(y_+)]=i\delta'(x_+-y_+)\,.
\ee
Its Hamiltonian is inherited from the scalar theory, which decouples into two chiralities (as well as any component of the stress tensor). Taking $j$ to be one of these chiralities, one obtains $H=\frac{1}{2}\int\dd x_+\ j^2$. Thus, gaussianity again implies that the whole theory is encoded in \eqref{eq:j_corr} and \eqref{eq:j_comm}.

The massless scalar field on the half-line is actually equivalent to the theory of the chiral current via the following identifications at $t=0$:
\begin{equation}
    \phi(x) = \frac{1}{\sqrt{2}} \int_{-x}^{x}  \dd y\ j(y)\,, \quad
    \pi(x) = \frac{1}{\sqrt{2}} \left( j(x) - j(-x) \right) ,
\label{eq:defhalflinej}
\end{equation}
where $x>0$. Note that at $t=0$ one has $x_+=x$, so we can safely omit the subindex in what follows. One must bear in mind, however, that the operator $j(x)$ is defined on a null-line. Using the equations \eqref{eq:j_corr} and \eqref{eq:j_comm}, one can check that \eqref{eq:halflinescalardef} holds for the fields defined in \eqref{eq:defhalflinej}. Conversely, with the reciprocal transformation
\begin{equation}
    j(x) = \frac{1}{\sqrt{2}} \left( \d_{x} \phi(x) + \pi(x) \right) , \quad
    j(-x) = \frac{1}{\sqrt{2}} \left( \d_{x} \phi(x) - \pi(x) \right) ,
\label{eq:defjhalfline}
\end{equation}
one can check, using the equations \eqref{eq:halflinescalardef}, that \eqref{eq:j_corr} and \eqref{eq:j_comm} hold for the currents defined in \eqref{eq:defjhalfline}.

\subsection{Identifying the algebras}

Now we will use the identifications \eqref{eq:defhalflinej} and \eqref{eq:defjhalfline} in order to relate the local algebras of both theories. We will consider three types of regions on the half-line: first we will deal with intervals unattached to the origin, then we will consider the attached case, and finally we will focus on the complementary region of the former one. This last type of region is crucial for the computation of the RMI in the following section. 

\begin{figure}[h!]
    \centering
    \includegraphics[width=0.7\linewidth]{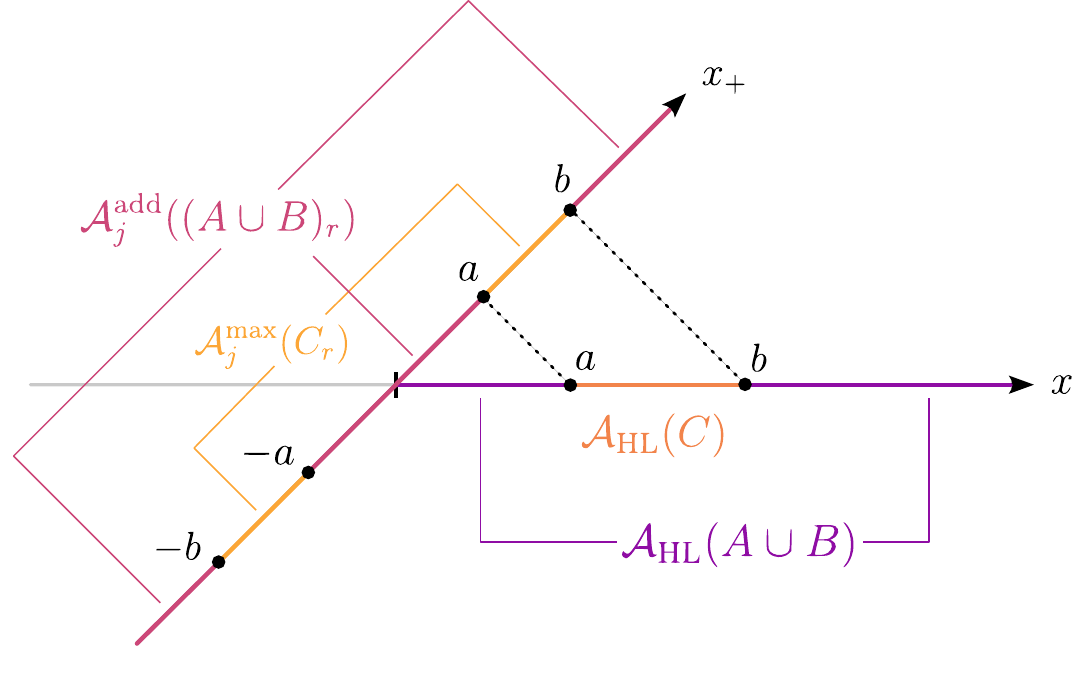}
    \caption{Geometric setup of some regions employed on the main text, as well as the identification between their corresponding algebras.}
    \label{fig:intervals}
\end{figure}

Consider first the interval $C = (a, b)$ with $0<a<b$. In the half-line theory, the algebra $\mathcal{A}_{\text{HL}}(C)$ associated with this interval is formed by all the polynomials of the operators $\phi$ and $\pi$ smeared with test functions of compact support inside $C$. The dual algebra on the chiral theory will turn out to be associated with the reflected interval on the null-line $C_r = -C \cup C$, where $-C = (-b, -a)$  as we show in Figure \ref{fig:intervals} (similar identifications have been used in theories with fermions \cite{Mintchev:2020uom}). In particular, it will be given by the so called maximal algebra over this region, which we denote $\mathcal{A}^{\text{max}}_{j}(C_r)$.
Maximal algebras arise in theories where Haag duality does not hold and there is an ambiguity in the assignment of algebras to regions \cite{Casini:2019kex,Casini:2021zgr}. One simple possibility is to take the algebra formed by all the polynomials of the current $j$ smeared with test functions of compact support inside the region, we will refer to this choice as the additive algebra $\mathcal{A}^{\text{add}}_{j}(C_r)$. In turn, the maximal algebra for some region $R$ is defined as the commutant of the additive algebra of the complementary region:
\be
{\cal A}_j^{\textrm{max}}(R)=({\cal A}^{\textrm{add}}_j(R'))'\,.
\ee
Here for an algebra ${\cal A}$ the commutant algebra is denoted ${\cal A}'$, and for a region $R$ the complementary region is called $R'$. Haag duality is the case in which ${\cal A}_j^{\textrm{max}}(R)={\cal A}^{\textrm{add}}_j(R)$, but this does not always hold. 
  Maximal algebras obviously contain the additive ones, but they also contain extended operators, usually called non-local operators. For instance, choosing a function $f$ such that it is equal to $1$ over $(-a,a)$ and has compact support inside $(-b,b)$, then
\begin{equation}\label{eq:j(f)}
    j(f) = \int \dd x \ f(x) j(x)\, ,
\end{equation}
is a non-local operator in $\mc A_j^{\text{max}}(C_r)$. We can show that this non-local operator commutes with the additive algebra of the complementary region $C_r'=(-\infty,-b)\cup(-a,a)\cup(b,\infty)$ by taking another function $g$ with compact support in $C_r'$, giving rise to an operator $j(g) \in \mathcal{A}^{\text{add}}_j(C_r')$. It turns out that
\begin{equation}
    \left[j(f), j(g) \right] 
    = i \int \dd x\, \dd y\ f(x) g(y) \delta^{\prime}(x-y)
    = i \int \dd y\ f^{\prime}(y) g(y) = 0 \ ,
\end{equation}
where the integral is zero because $f^{\prime}(y) = 0$ for $y\in C_r'$. However, note that the operator $j(f)$ cannot be generated with local operators in $C_r$, since $f(x)=1$ for $x\in(-a, a) \subset C_r'$. Then, we have $j(f) \in \mathcal{A}^{\text{add}}_j(C_r')^{\prime}$ and $j(f) \notin \mathcal{A}^{\text{add}}_j(C_r)$. Nonetheless, we remark that this non-local operator can be generated locally in the additive algebra associated with the interval $(-b, b)$, because $f$ has compact support there.

Next, we will prove the claim that
\begin{equation}
    \mathcal{A}_{\text{HL}}(C) = \mathcal{A}^{\text{max}}_j(C_r) \, .
\label{eq:Algebranoattached}
\end{equation}
To show that $\mathcal{A}^{\text{max}}_j(C_r) \subset \mathcal{A}_{\text{HL}}(C)$, we use \eqref{eq:defjhalfline} to write $j(g)$ for some function $g$ as
\begin{equation}
\small
\begin{split}
    \int_{-\infty}^{\infty}\dd x\ g(x) j(x) 
    &= \frac{1}{\sqrt{2}} \left[ \int_{0}^{\infty} \dd x \of{g(x) + g(-x)} \d_{x} \phi(x) + \int_{0}^{\infty}\dd x  \of{g(x) - g(-x)} \pi(x) \right] \\
    &= \frac{1}{\sqrt{2}} \left[  \int_{0}^{\infty}\dd x \of{g^{\prime}(-x) - g^{\prime}(x)} \phi(x)  + \int_{0}^{\infty}\dd x \of{g(x) - g(-x)} \pi(x) \right]\ .
\end{split}
\end{equation}
If $g$ has compact support inside $C_r$ then $j(g) \in \mathcal{A}^{\text{max}}_j(C_r)$. Also, note that $g(x) - g(-x)$ and $g^{\prime}(-x) - g^{\prime}(x)$ have compact support on $C$ as functions of $x>0$, therefore $j(g) \in \mathcal{A}_{\text{HL}}(C)$. In the maximal algebra we also have the non-local operators, which correspond to taking $g$ constant in $(-a, a)$ and zero outside $(-b, b)$. Again, in this case we have that $g(x) - g(-x)$ and $g^{\prime}(-x) - g^{\prime}(x)$ are supported in $C$ and thus $j(g) \in \mathcal{A}_{\text{HL}}(C)$. Therefore, we conclude that the inclusion $\mathcal{A}^{\text{max}}_j(C_r) \subset \mathcal{A}_{\text{HL}}(C)$ holds.

To prove the other inclusion, we use \eqref{eq:defhalflinej} to write $\pi(g)$ as
\begin{equation}
    \pi(g) = \frac{1}{\sqrt{2}} \int_{0}^{\infty}\dd x\  g(x) \of{ j(x) - j(-x) } 
    = \frac{1}{\sqrt{2}} \int_{-\infty}^{\infty} \dd x\ g_{-}(x) j(x)  \, ,
\end{equation}
where $g_{-}$ is the odd extension of $g$ to the full line. If $g$ is of compact support in $C$, we have that $\pi(g) \in \mathcal{A}_{\text{HL}}(C)$. Also, the function $g_{-}$ has compact support in $C_r$, and we have that $\pi(g) \in \mathcal{A}^{\text{max}}_j(C_r)$. 
If we now look at the operator $\phi(g)$ we have that
\begin{equation}
\begin{split}
    \phi(g) = \int_{0}^{\infty}\dd x\ g(x) \phi(x) 
    &=  \int_{0}^{\infty}\dd x\ g(x) \left( \int_{-x}^{x}\dd y \ j(y) \right)  \\
    &=  \int_{0}^{\infty} \dd x\ g(x) \left( \int_{-\infty}^{\infty}\dd y\ \chi_{(-x,x)}(y) j(y) \right) \dd x \\
    &=  \int_{-\infty}^{\infty} \dd y \  \widetilde{g}(y) j(y) \, ,
\end{split}
\end{equation}
where $\chi_{(-x,x)}(y)$ is the characteristic function of the interval $(-x,x)$ and
\begin{equation}
    \widetilde{g}(y) = \int_{0}^{\infty} \dd x\ \chi_{(-x,x)}(y) g(x) \, .
\end{equation}
If $g$ has compact support in $C$, we have that $\phi(g) \in \mathcal{A}_{\text{HL}}(C)$. It is easy to check that $\widetilde{g}$ has compact support in $(-b, b)$ and that is constant in $(-a, a)$. Therefore, by the discussion given below \eqref{eq:j(f)}, we have that $\phi(g) \in \mathcal{A}^{\text{max}}_j(C_r)$. This proves $\mathcal{A}_{\text{HL}}(C) \subset \mathcal{A}^{\text{max}}_j(C_r)$ and thus the equality \eqref{eq:Algebranoattached} is established.

Now we turn to the case of an interval attached to the origin. Taking $A=(0,a)$, its reflected partner $A_{r} = -A \cup A = (-a,a)$ is also a single interval on the null-line. The algebra of one interval in the theory of the chiral current satisfies Haag duality \cite{Buchholz:1990ew}. Therefore, there is no ambiguity in the choice of algebras in this case, thus the identification is
\begin{equation}
    \mathcal{A}_{\text{HL}}(A) = \mathcal{A}^{\text{max}}_{j}(A_{r}) = \mathcal{A}^{\text{add}}_{j}(A_{r}) \, .
\end{equation}
Finally, we focus on the type of region essential for the rest of this work. We will consider the algebra formed by the operators living in region $A \cup B$, where
$A = (0,a)$ and $B = (b, \infty)$, which is constructed as\footnote{Here the symbol $\vee$ means that $\mathcal{A} \vee \mathcal{B} = (\mathcal{A} \cup \mathcal{B})^{\prime \prime}$. It gives the smallest von Neumman algebra containing both $\mc A$ and $\mc B$. In our case, is the algebra formed by the multiplications, linear combinations, and limits of all the operators on both regions.}
\begin{equation}
    \mathcal{A}_{\text{HL}}(A \cup B ) = \mathcal{A}_{\text{HL}}(A) \vee \mathcal{A}_{\text{HL}}(B)\,.
\end{equation}
We have just shown that for intervals attached to the origin, the algebra on the half-line is equal to the algebra of a single interval on the null-line. This is the case for $A$, and in particular it is also the case for $B' = (0 , b)$. Now, since Haag duality holds for the algebra of single intervals attached to the origin on the half-line,\footnote{This can be intuitively understood from the fact that the modular conjugation $J$ associated to the algebra of an interval attached to the origin acts geometrically \cite{Huerta:2022tpq, Garbarz:2021rqu,ECKMANN19731}.} we have that
\begin{equation}
    \mathcal{A}_{\text{HL}}(B) = (\mathcal{A}_{\text{HL}}(B'))' = (\mathcal{A}^{\text{add}}_{j}((B')_{r}))' = \mathcal{A}^{\text{add}}_{j}(B_{r}) \, . 
\end{equation}
Here, we used that $(B^{\prime})_{r} = (B_{r})^{\prime}$ where the complement in the left-hand side refers to a complement in the half-line and in the right-hand side on the null-line. With this, we conclude that
\begin{equation}
    \mathcal{A}_{\text{HL}}(A \cup B ) 
    = \mathcal{A}_{\text{HL}}(A) \vee \mathcal{A}_{\text{HL}}(B)
    = \mathcal{A}^{\text{add}}_{j}(A_{r}) \vee \mathcal{A}^{\text{add}}_{j}(B_{r})
    = \mathcal{A}_{j}^{\text{add}}((A \cup B)_{r}) \, .
\label{eq:algebraABinteres}
\end{equation}
So the algebra of the region $A\cup B$ on the half-line is identified with the additive algebra of the chiral current for its reflected region. As stated earlier, this fact will be extremely relevant for computing RMIs in the next section.

To end this section, we would like to remark an interesting corollary that can be derived from \eqref{eq:algebraABinteres}. Noting that $(A \cup B)' = (a,b)=C$, one has
\begin{equation}
    (\mathcal{A}_{\text{HL}}(C'))'
    = (\mathcal{A}_{j}^{\text{add}}((A \cup B)_{r}))'
    = \mathcal{A}_{j}^{\text{max}}((A \cup B)_{r}')
    = \mathcal{A}_{\text{HL}}(C) \, .
\end{equation}
Then we show that the scalar field on the half-line also satisfies Haag duality for intervals unattached to the origin.

\section{Rényi mutual information}\label{sec:3}

\begin{figure}[t]
    \centering
    \includegraphics[width=0.9\linewidth]{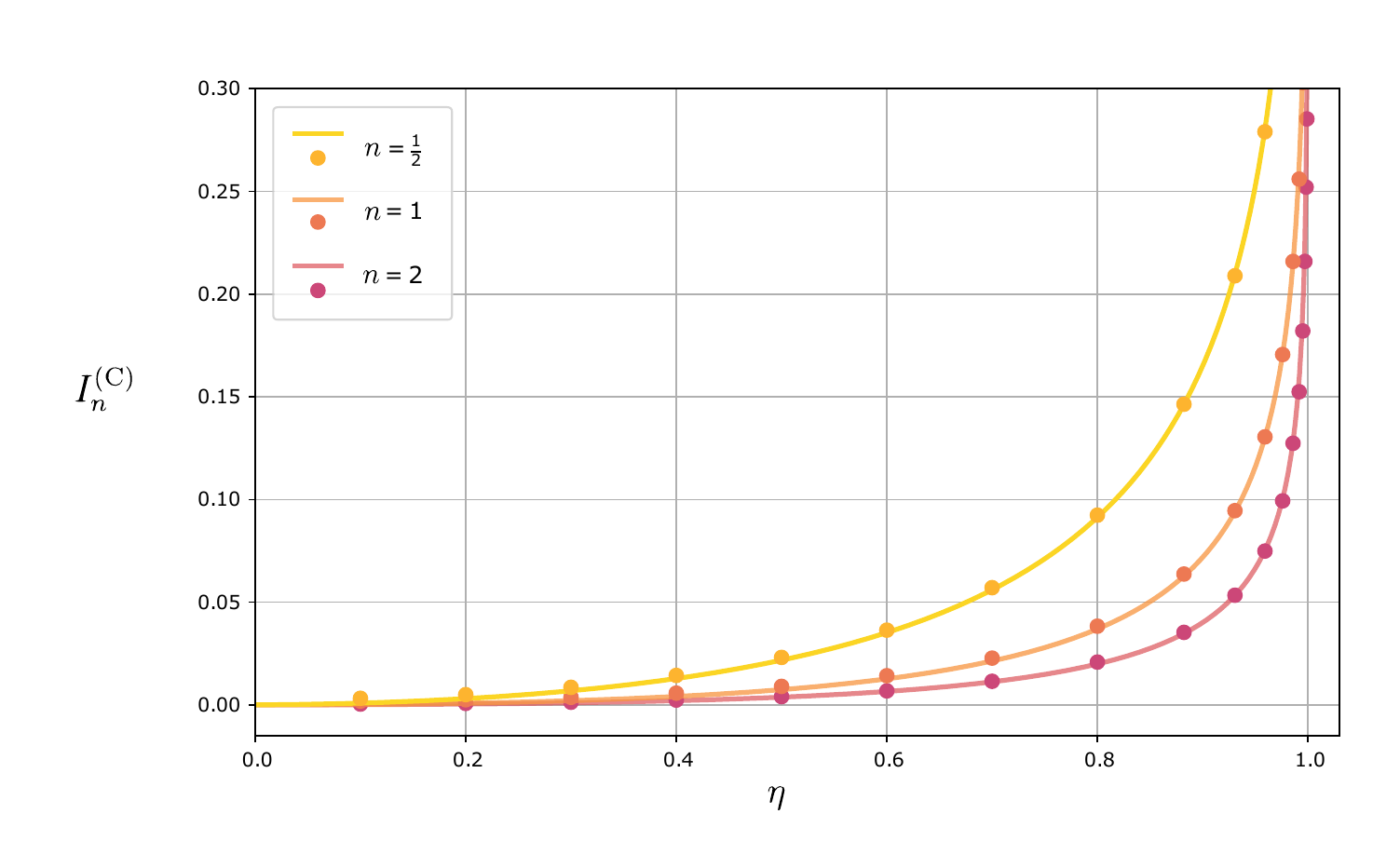}
    \caption{In circles we show the numerical calculation of the $n$-th RMIs for the scalar on the half-line performed on a lattice for different values of the cross-ratio $\eta$ and for $n=0.5,1,2$ (from top to bottom). In solid lines we show  the corresponding analytical results.}
    \label{fig:numerics}
\end{figure}

In this section, we will study the RMIs for the massless scalar field on the half-line in $d=2$, satisfying Dirichlet boundary conditions at the origin, for the two intervals $A=(0,a)$ and $B=(b,\infty)$, shown in Figure \ref{fig:intervals}. An expression for these RMIs for integer $n$ was calculated in \cite{Estienne:2023ekf} by taking the decompactification limit of the results from the compactified scalar. Here, we present an alternative approach that applies to generic $n$. Once the algebra identifications \eqref{eq:algebraABinteres} between the scalar field on the half-line and the chiral current are established, the computation of the $n$-th RMI becomes straightforward. This is because $(A\cup B)_r$ is given by the union of two intervals (once we compactify the null-line into a circle), and the $n$-th RMI of the additive algebra of such configuration in the chiral current theory was derived in \cite{Arias:2018tmw}. Then, the $n$-th RMI for the massless scalar field in $d=2$ with Dirichlet boundary condition at the origin is equal to that of the chiral current for the corresponding two-interval configuration 
\be\label{eq:RMI_result}
I^{(0)}_{n}(A,B) = I^{(\text{C})}_n(A_{r},B_{r})=-\frac{n+1}{12n}\log(1-\eta)+U_n(\eta)\,,
\ee
where the dependence on the cross-ratio
\be\label{eq:eta}
\eta=\frac{4ab}{(a+b)^2}\,,
\ee
comes from conformal invariance, the function $U_n$ is given by
\be
U_n(\eta)=\frac{i n}{2(n-1)}\int_0^\infty\dd s\of{\coth(n\pi s)-\coth(\pi s)}\log \left( \frac{F(s,\eta)}{F(-s,\eta)} \right),
\label{eq:Unanalytic}
\ee
and $F(s,\eta)$ is the following hypergeometric function
\be
F(s,\eta)={}_2F_1(1 + i s, -i s; 1; \eta)\,.
\ee
In particular, taking the limit $n\to1$ on \eqref{eq:RMI_result} we also arrive at the MI,\footnote{One can also arrive to this expression through a more lengthy derivation. It consists in identifying the entropy of $A\cup B$ on the algebra for the scalar on the half-line with the entropy of $(A\cup B)_r$ on the additive algebra of the chiral current, again using \eqref{eq:algebraABinteres}. The latter EE is obtained from the MI of the chiral current by subtracting the entropies of $A_r$ and $B_r$ for this theory. Then, adding the entropies of $A$ and $B$ for the scalar on the half-line yields \eqref{eq:MI}. All the EEs employed are well-known in the literature: they are simply given by the logarithm of the size of the interval, plus a boundary term in the case of the theory on the half-line \cite{Calabrese:2009qy}. Remarkably, the boundary term in the EE of $A$ and $B$ precisely cancels the contribution coming from the difference in the size between the intervals $A$,$B$ and $A_r$,$B_r$.}
\be\label{eq:MI}
I^{(0)}(A,B) = I^{(\text{C})}(A_{r},B_{r}) = -\frac{1}{6}\log(1-\eta)+U(\eta)\,,
\ee
where
\be
U(\eta) = -\frac{i \pi}{2} \int_0^{+\infty} \mathrm{d}s \, \frac{s}{\sinh^2(\pi s)} \log \left( \frac{F(s,\eta)}{F(-s,\eta)} \right).
\ee
This shows that (for $n=1$) the corresponding expressions obtained in \cite{Estienne:2023ekf} for the scalar in the half-line and the current in \cite{Arias:2018tmw} coincide.
We have checked these results numerically by evaluating the entropies $S^{(0)}_n(A)$, $S^{(0)}_n(B)$, and $S^{(0)}_n(A\cup B)$ on a lattice using standard computation methods in terms of correlators \cite{Casini:2009sr} (see also \cite{Casini:2015dsg} for explicit expressions for the two-point functions on the lattice). Combining these entropies we were able to obtain $I^{(0)}_n(A,B)$ for different values of $a,b$ and $n$. These results, together with the analytic expressions, are shown in Figure \ref{fig:numerics} where we can see perfect agreement between the curves.

Coming back to the higher dimensional $d=4$ problem of the difference of RMI between two scalars and a Maxwell field, the results of this section and the preceding one can be sumarized in the formula 
\be
I_n^{(\text{M})}(\eta)=2\of{I_n^{(\text{S})}(\eta)-I_n^{(\text{C})}(\eta)}=2\, I_n^{(\text{S})}(\eta)+\frac{n+1}{6n}\log(1-\eta)-2 \, U_n(\eta) \,. \label{dfg}
\ee

\subsection{Short and long distance expansions}
 
The long and short distance limits of our results can be further studied. Let us first discuss the short distance limit. Taking $b\sim a$ in \eqref{eq:eta}, which implies $\eta\sim1$, the second term in the right hand side of (\ref{dfg}) provides a logarithmic divergence coming from the separation distance 
\be
\epsilon=b-a\ll a\,, \quad 1-\eta\simeq \epsilon^2/(4 a)^2\,.
\ee
On the other hand, this limit for the function $U_{n}$ has been carefully studied in \cite{Arias:2018tmw} yielding a subleading divergent contribution. Combining both we get
\be\label{eq:shor_distance}
 2\, I_n^{(\text{S})}-I_n^{(\text{M})}\simeq -\frac{n+1}{3n}\log(\epsilon/a)-\log\of{-\log(\epsilon/a)}\qquad\text{for}\qquad \eta\to 1\,\,\, (\epsilon\rightarrow 0)\,.
\ee
This behavior is to be expected since in this limit the RMI operates as a geometrical regulator for (twice) the Rényi EE. Namely, the first contribution precisely gives twice the universal term for the Rényi entropy of an interval in a CFT with central charge $c=1$, while the $\log\of{-\log(\epsilon/a)}$ contribution is associated with the infrared divergence of the massless scalar in $d=2$ \cite{Casini:2009sr}. Notice this last contribution has ``flat spectrum'', namely, it is independent of $n$.

This limit can be used to find a regularization-independent value for the logarithmic coefficient of the EE of the Maxwell field $I^{(\text {M})}_{\textrm{log}}/2\simeq S^{(\text {M})}_{\textrm{log}}\simeq a_{\text{M}}\, \log(\epsilon/a)$. 
Taking into account that the logarithmic coefficient for the scalar in $d=4$ is $a_{\text{S}}=1/90$ \cite{casini:2010kt}, using \eqref{eq:shor_distance} for $n=1$ one gets $a_{\text{M}}=16/45$. This was previously done in \cite{Casini:2015dsg}, obtaining a value that coincides with results found using thermodynamic methods \cite{Dowker:2010bu}. Here we arrived at the same result via the limit of the exact analytic expression of the MI difference.

Now we turn to the long distance limit. This expansion for the MI provides information about the operatorial content of the theory. Each Rényi twist for the two regions is expanded in local operators for the replicated theory. The leading contribution corresponds to two-copy operators. In this limit it is known to be proportional to $L^{-4\Delta}\sim \eta^{2 \Delta}$, where $L$ is the separation distance between the regions and $\Delta$ is the lowest dimension among all the primary operators of the CFT \cite{Cardy:2013nua,Calabrese:2009ez,Calabrese:2010he}.  The subsequent subleading powers of $L$ in the expansion come from the contribution of higher dimensional operators on a replicated theory, since this result relies on computing the $n$-th RMI using the replica trick, which for the MI have to be analytically continued to $n\to1$. In this limit, expansions for $I^{(\text{M})}$ and $I^{(\text{S})}$ in powers of $\eta$ are available. For the Maxwell case, the leading coefficient has been explicitly computed in \cite{Casini:2021raa}, giving
\be\label{eq:Maxwell_MI}
I^{(\text{M})}= \frac{1}{210}\eta^4+\mc O\of{\eta^5}.
\ee
The quartic power comes from the fact that the operator with lowest scaling dimension on this theory is the field strength tensor $F_{\mu\nu}$, with $\Delta=2$. Meanwhile, in the scalar case, even some subleading terms have been computed \cite{Chen:2016mya}. The contribution up to $\eta^4$ is known to be 
\be\label{eq:scalar_MI}
I^{(\text{S})}=\frac{1}{60}\eta^2+\frac{1}{70}\eta^3+\frac{13}{840}\eta^4+\mc O\of{\eta^5}.
\ee
The primary with lowest dimension in this theory is the field itself with $\Delta=1$, explaining the leading power in the expansion. The subleading powers come from the contribution of operators of higher dimension in the $n$-th replicated theory in the limit $n\to1$, i.e., powers and derivatives of the scalar. 

Replacing \eqref{eq:Maxwell_MI} and \eqref{eq:scalar_MI} into \eqref{dfg}, we should in principle obtain the long separation expansion of our result for the scalar on the half-line up to quartic order. This combination gives
\be
I^{(\text{S})}-\frac{1}{2}\,I^{(\text{M})}=\frac{1}{60}\eta^2 
    + \frac{1}{70}\eta^3 
    + \frac{11}{840}\eta^4
    + \mathcal{O}\of{\eta^{5}} ,
\ee
which precisely matches the series expansion of \eqref{eq:RMI_result}, providing a crosscheck of our computation. The power of the leading term is consistent with the fact that the operator with lowest dimension on the theory of the chiral current is the one with $\Delta=1$. We will carefully study this expansion in the next section, as well as the corresponding to $n\neq1$, since their behaviors are quite interesting. The general result for the expansion of \eqref{eq:RMI_result} up to quartic order is
\begin{align}\label{eq:long_limit_RMI}
I^{(\text{C})}_{n}& = 
    \frac{(1+n)(1+n^2)}{240n^3} \eta^2 
    + \frac{(1+n)(4n^2-1)(5+4n^2)}{3780n^5} \eta^3 \, \notag \\
    &\qquad\quad + \frac{(1+n)(231-529n^2+941n^4+941n^6)}{241920n^7} \eta^4 \, 
    + \mathcal{O}\of{\eta^{5}} .
\end{align}
A large separation expansion for the RMI of the scalar in $d=4$ is also available in \cite{Chen:2016mya}, found to be 
\begin{align}\label{eq:scalar_long}
I_n^{(\text{S})}&=\frac{(1+n)(1+n^2)}{240n^3}\eta^2+\frac{(1+n)(4n^2-1)(5+4n^2)}{3780n^5}\eta^3\,\notag\\
&\qquad\quad +\frac{(1+n)(7-13n^2+29n^4+29n^6)}{6720n^7}\eta^4+\mc O\of{\eta^5}.
\end{align}
Notice the first two terms in (\ref{eq:long_limit_RMI}) and (\ref{eq:scalar_long}) agree, as it should, because the Maxwell field cannot give contributions to this order.

Meanwhile, to the best of our knowledge, not even the leading term for the $n$-th RMI of the Maxwell field for $n\neq 1$ has been computed. To end this section, we use our result to find this contribution. Combining \eqref{eq:scalar_long} with \eqref{eq:long_limit_RMI} into \eqref{dfg}, we obtain
\be
I_n^{(\text{M})}=\frac{(1+n)(21+61n^2+103n^4+103n^6)}{120960n^7}\eta^4+\mc O\of{\eta^5},
\ee
which yields \eqref{eq:Maxwell_MI} after setting $n=1$.

\section{Long distance expansion asymptotics}
\label{sec:33}

\begin{figure}[h!]
    \centering
    \includegraphics[width=0.9\linewidth]{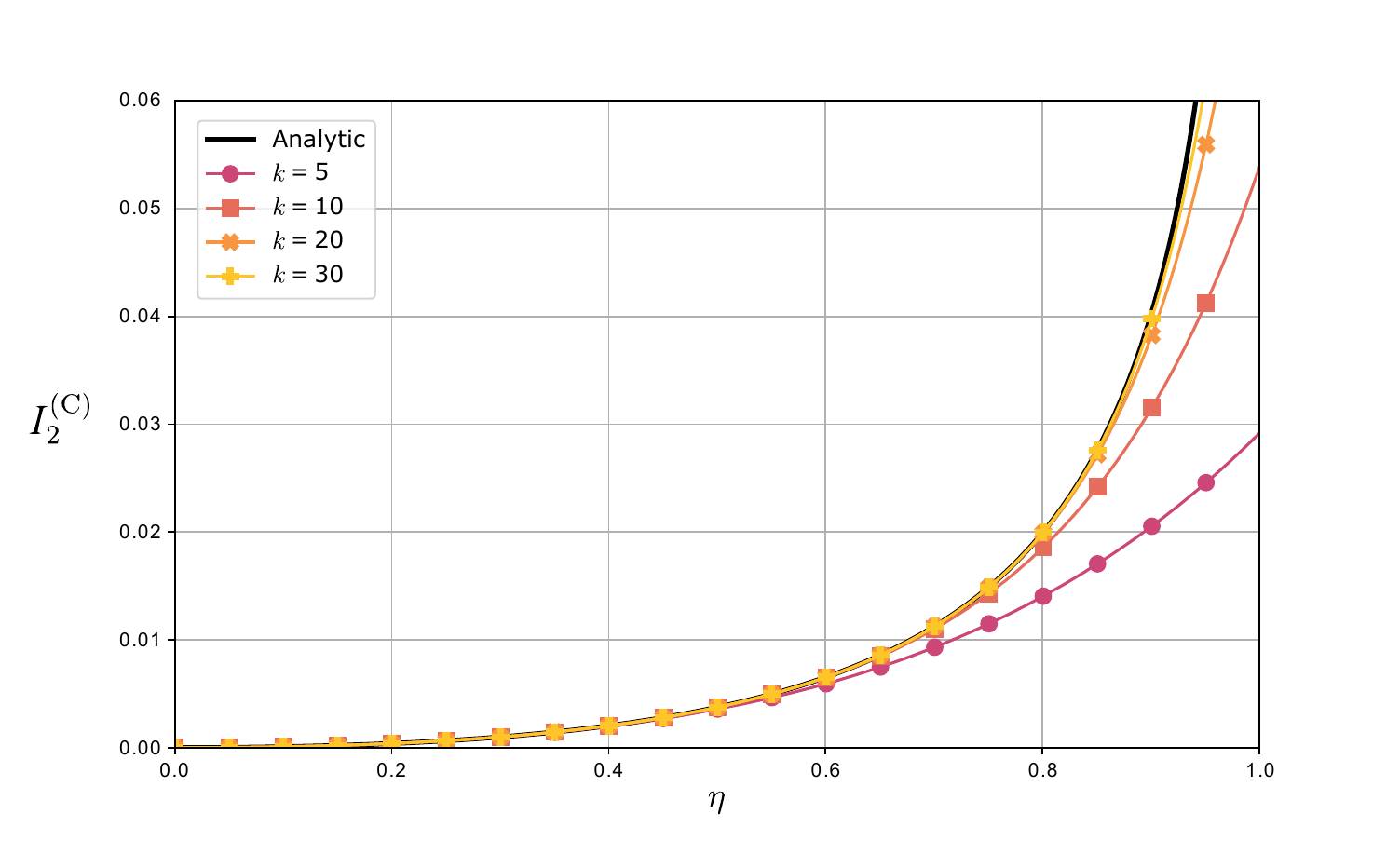}
    \caption{Comparison between the analytic expression for the RMI \eqref{eq:RMI_result} with $n=2$ and it's long distance expansion for different orders $k$ of the expansion. The same behavior is observed for other integer values of $n\neq1$.}
    \label{fig:nrmiseries}
\end{figure}

In this section we will study the long distance expansion \eqref{eq:long_limit_RMI} of \eqref{eq:RMI_result} and (\ref{dfg}) in greater detail. An easy way to obtain such expression is to expand the integrand in $U_n$, then integrate in $s$ and finally combine this with the series expansion for the logarithm. For the first step, we write
\begin{equation}
   \frac{i}{2} \log \left( \frac{F(s, \eta)}{F(-s, \eta)} \right)  = M_{1}(s)\, \eta + M_{2}(s) \frac{\eta^{2}}{2!} + \cdots \ ,
\label{eq:exploghyper}
\end{equation}
where there is no zeroth order contribution since $F(s, 0) = 1$, and the coefficients $M_{k}(s)$ are odd polynomials of order $2k - 1$, because the function \eqref{eq:exploghyper} and all its derivatives are odd in $s$. Then, using that
\begin{equation}
    \int_{0}^{\infty}\dd s \left( \coth(n \pi s) - \coth(\pi s) \right) s^{k} 
    = \frac{2 k!  \, \zeta(k + 1) }{(2 \pi)^{k + 1}} \left( \frac{1}{n^{k + 1}} - 1 \right),
\label{eq:integralcoth}
\end{equation}
we can explicitly integrate the series on the right-hand side of \eqref{eq:exploghyper} against the factor $(\coth(n\pi s)-\coth(\pi s))$ to any order. With this, we can write the series expansion of $U_n$ and therefore the long distance limit of the RMI. Up to quartic order, this leads to \eqref{eq:long_limit_RMI}. In Figure \ref{fig:nrmiseries} we show the long distance expansion for $I^{(\text{C})}_2(\eta)$ for higher orders. We can clearly see that this power series converges to the analytic result, uniformly in each interval $(0,y)$, $y<1$, and the same behavior is observed for higher integer values of $n$.

\begin{figure}[h!]
    \centering
    \includegraphics[width=0.9\linewidth]{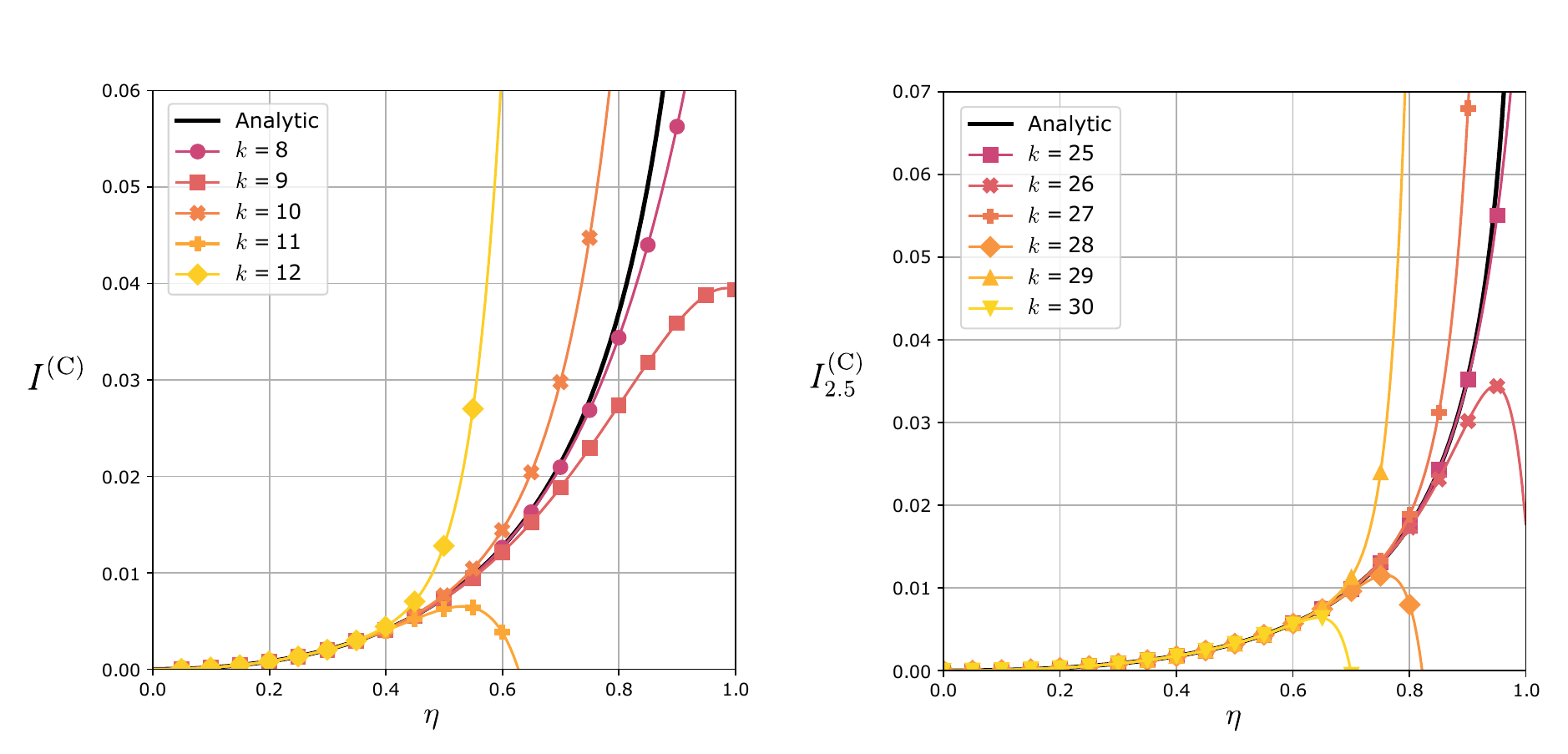}
    \caption{On the left we show the comparison between the analytic expression for the MI \eqref{eq:MI} and it's long distance expansion up to the $k$-th order, for different values of $k$. We can see that the convergence is spoiled with increasing $k$. The same behavior is observed for the RMIs \eqref{eq:RMI_result} with $n\not\in\mathbb{Z}$, as we show on the right panel for $n=2.5$}
    \label{fig:rmiseries}
\end{figure}

However, something rather puzzling happens when $n$ hits a non-integer value as well as  when $n=1$. For instance, with the same procedure we can get the large separation limit of the MI at any order in $\eta$. Again up to $\eta^4$, this is
\be\label{eq:long_limit_MI}
    I^{(\text{C})}(A,B) = \frac{1}{60}\eta^2 
    + \frac{1}{70}\eta^3 
    + \frac{11}{840}\eta^4
    + \mathcal{O}\of{\eta^{5}},
\ee
which coincides with the combination $I^{(\text{S})}-I^{(\text{M})}/2$ as stated in the previous section. Note that the coefficients in \eqref{eq:long_limit_MI} can also be computed replacing $n=1$ in \eqref{eq:long_limit_RMI}. We show higher orders of the expansion in Figure \ref{fig:rmiseries}, where surprisingly we can see that there is no improvement in the converge to the analytical result as the order is increased. We also show the long distance expansion \eqref{eq:long_limit_RMI} for $n=2.5$, where again increasing the order spoils the convergence. 

\begin{figure}[h!]
    \centering
    \includegraphics[width=0.9\linewidth]{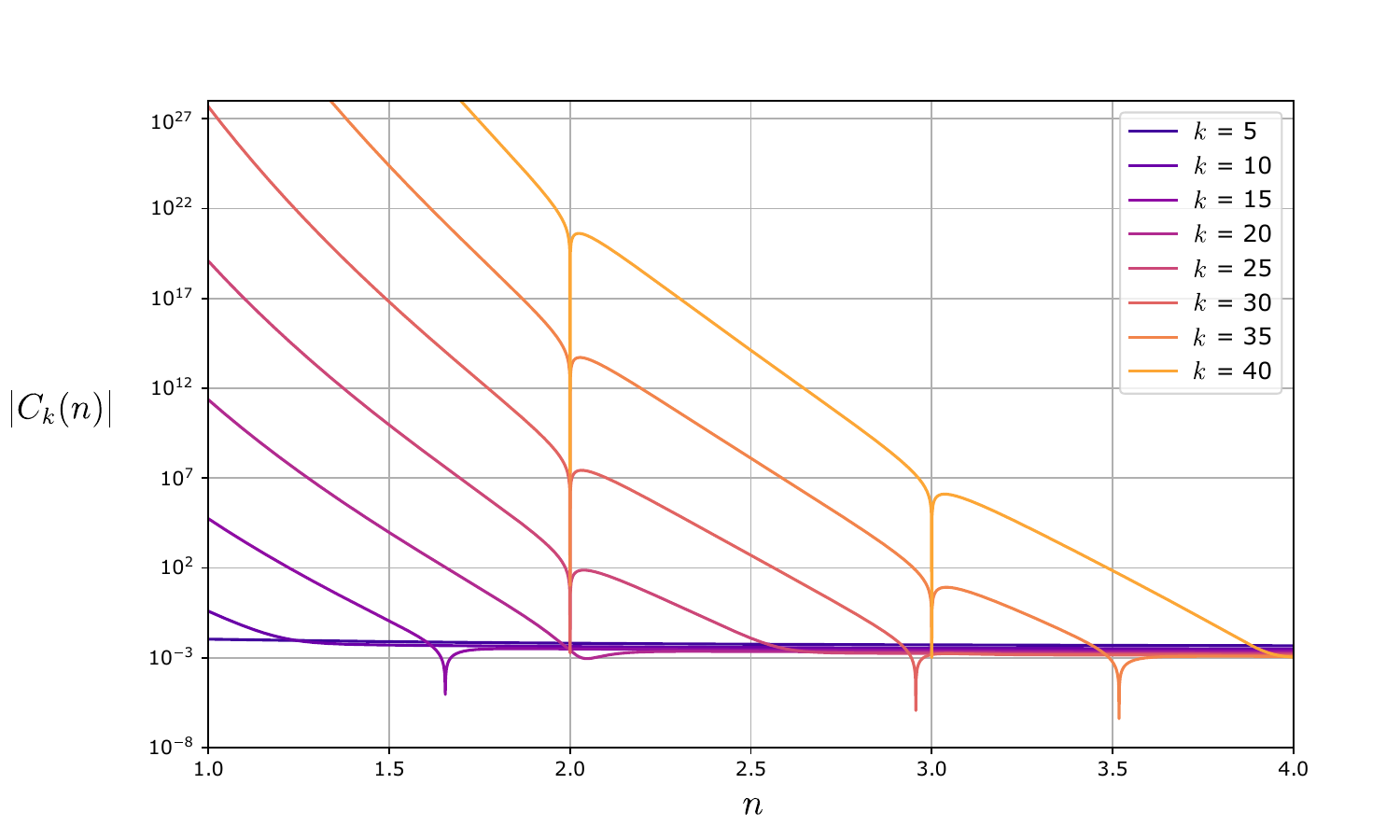}
    \caption{
        Plot of the absolute value of the $k$-th coefficient in the long distance expansion of $I^{(\text{C})}_{n}(\eta)$ for some values of $k$ between 5 and 40 (from bottom to top), as a function of $n$. Note the logarithmic scale for the vertical axes. 
    }
    \label{fig:plot_coeff_n}
\end{figure}

We can visualize the failure of the convergence in the long distance expansion of the RMI $I^{(\text{C})}_n(\eta) \sim \sum_{k} C_{k}(n)\,\eta^{k}$ for arbitrary $n \geq1$ in Figure \ref{fig:plot_coeff_n}. There, we plot the absolute value of the coefficients $C_{k}(n)$ for different values of $k$ as a function of $n$. In this plot we can explicitly see $| C_k(n) |$ exponentially increases with $k$, except for integers values of $n$, where the coefficients collapse and do not grow exponentially with $k$.

An explanation for this peculiar behavior is that the power series expansion \eqref{eq:exploghyper} becomes somewhat ill-defined under the integral sign in \eqref{eq:Unanalytic}. To see this, let us first analyze the radius of convergence of the power expansion \eqref{eq:exploghyper} using the root test. Consider the limit
\begin{equation}
\begin{split}
    \limsup_{k \rightarrow \infty} \sqrt[k]{\abs{\frac{M_{k}(s)\,\eta^{k}}{k!}}}
    & = \limsup_{k \rightarrow \infty} \sqrt[k]{\abs{\frac{M_{k}(s) }{s^{2k-1} k!}}} s^{2-1/k} \abs{\eta} 
\equiv c(s) s^{2} \abs{\eta}\, ,
\end{split}
\label{eq:root_test}
\end{equation}
where we have defined
\begin{equation}\label{eq:c(s)}
    c(s) = \limsup_{k \rightarrow \infty} \sqrt[k]{\abs{\frac{M_{k}(s) }{s^{2k-1} k!}}} \, .
\end{equation}
In order for \eqref{eq:root_test} to be less than 1, we need
\be\label{eq:radconvergences}
|\eta|<\frac{1}{c(s)s^2}\,,
\ee
which sets the radius of convergence, crucially depending on $s$. For large $s$, since the $M_k(s)$ are polynomials of degree $s^{2k-1}$, the ratio $M_k/s^{2k-1}$ becomes a constant that we can compute for different (finite) values of $k$.  Using this on \eqref{eq:c(s)}, we have numerically checked that $c(s\to\infty) \sim c\equiv 0.69166$, and thus the radius of convergence becomes zero in this limit.\footnote{
Another way to obtain the radius of convergence in \eqref{eq:radconvergences} is by noticing that the hypergeometric function ${}_2F_1(1 + i s, -i s; 1; z)$ vanishes at a point on the complex plane that, for large $s$, can be approximated by $z \approx - (c s^{2})^{-1} - i (c s^{3})^{-1}$. Then, the logarithm in $U_n$ diverges near $\eta\sim0$ for $s\to\infty$.
} Since the integration in \eqref{eq:Unanalytic} extends to infinity, and there we have zero radius of convergence, we see that performing the expansion inside the integral is not allowed. Interestingly enough, we have also performed a series expansion of \eqref{eq:RMI_result} numerically after integration, obtaining the same coefficients $C_k(n)$. This tells us that, generically, the $n$-th RMI actually is not expected to admit a proper Taylor expansion at $\eta=0$, and the series in powers of $\eta$ is rather of the asymptotic type.\footnote{
The function $S(\epsilon) = \int_{0}^{\infty} \dd t\ e^{-t} (1 + \epsilon t)^{-1}$
provides a simpler example where the same issue arises when attempting a Taylor expansion around $ \epsilon = 0 $. This example, along with others, is lengthy discussed in \cite{boyd_devils_1999}.
}

\begin{figure}[t]
    \centering
    \includegraphics[width=0.5\linewidth]{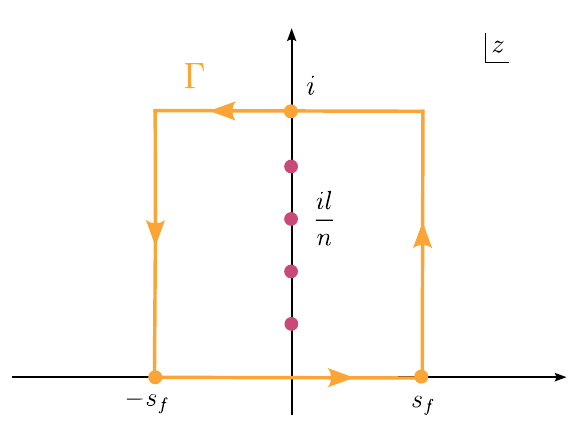}
    \caption{
        Representation of the contour of integration $\Gamma$ over the complex plane for the integral \eqref{eq:contourintegral}, as well as the poles of the integrand on the imaginary axis.
    }
    \label{fig:contourintegration}
\end{figure}

However, the story is different for integer $n \geq 2$ because the integral in \eqref{eq:Unanalytic} can be represented as a finite sum \cite{Estienne:2023ekf}.  Consider the following contour integral over the complex plane
\begin{equation}
    \oint_{\Gamma} \dd z \of{\coth(n\pi z)-\coth(\pi z)} \log \left( F(z, \eta) \right) ,
\label{eq:contourintegral}
\end{equation}
where $\Gamma$ is the a rectangle with vertices $\{ -s_{f}, s_{f}, s_{f} + i, -s_{f} + i \}$ drawn in Figure \ref{fig:contourintegration}. The integrand is holomorphic inside this contour except for the poles at $z = il/n$, that come from the factor $\coth(n \pi z)$ for $l=1,\dots n-1$; and regular at the points $s=0$ and $s=i$, because $F(0,\eta) = F(i, \eta) = 1$.
The paths of the contour parallel to the imaginary axis vanish when we take $s_{f} \rightarrow \infty$ because of the exponential damping of $\of{\coth(n\pi z)-\coth(\pi z)}$. Therefore, we can consider only the paths of the contour parallel to the real axis. These give
\begin{equation}
\begin{split}
    &\oint_{\Gamma} \dd z \of{\coth(n\pi z)-\coth(\pi z)} \log \left( F(z, \eta) \right) \\
    &\qquad = 2 \int_{0}^{\infty} \dd s \of{\coth(n\pi s)-\coth(\pi s)} \log \left( \frac{F(s, \eta)}{F(-s, \eta)} \right),
\end{split}
\end{equation}

\noindent where we have used that $F(-s+i,\eta) = F(s,\eta)$ and that $\coth(n\pi s)-\coth(\pi s)$ is odd in $s$ and, for integer $n$, it is periodic in the imaginary axis with period $s \sim s+i$. 
Therefore, when we have integer $n \geq 2$ we can identify the integral in \eqref{eq:Unanalytic} with the contour integral \eqref{eq:contourintegral}. Finally, since $\Gamma$ is a closed contour, we can compute the integral using the residue theorem. This gives
\begin{equation}
    \int_{0}^{\infty} \dd s \of{\coth(n\pi s)-\coth(\pi s)} \log \left( \frac{F(s, \eta)}{F(-s, \eta)} \right) 
    = \frac{i}{n} \sum_{l=1}^{n-1} \log \left[ F \left( \frac{il}{n}, \eta \right) \right] \ ,
\end{equation}
and replacing this in \eqref{eq:Unanalytic}, we obtain
\begin{equation}
    U_n(\eta) = - \frac{1}{2(n-1)} \sum_{l=1}^{n-1} \log \left[ F \of{ \frac{il}{n}, \eta } \right] ,
\label{eq:Unsuma}
\end{equation}
which is only valid for $n\in\mathbb{Z}_{\geq2}$. This matches with previous results for the scalar on the half-line \cite{Estienne:2023ekf}. Since $F\!\of{\frac{il}{n}, \eta}$ is analytic at $\eta = 0$, where it takes the value $1$, and its logarithm is also analytic at that point, it follows that \eqref{eq:Unsuma} is analytic at $\eta = 0$. Therefore, we conclude that the $n$-th RMI admits a Taylor expansion around $\eta = 0$ for integer $n \geq 2$, and performing this expansion we recover \eqref{eq:long_limit_RMI}.

We remark that the finite radius of convergence of the RMI for these values of $n$ could had been anticipated from the fact that it is given by the expectation values of replica twist operators \cite{Estienne:2023ekf,Cardy:2007mb,long2016co}. Correlators of localized operators in a CFT have a OPE expansion with a finite radius of convergence. However, for non integer $n$ and $n=1$ the expansions of the entropies is not an expansion of operator correlators.

\section{Outlook and future directions}\label{sec:4}

In this work we gave an explicit expression for the $n$-th Rényi mutual information (RMI) of a massless scalar field living on the half-line. We arrived at this result by  identifying the operator algebras of this theory to the corresponding ones for a chiral current living on the whole null-line. We saw that the algebras on the half-line for certain configuration of intervals could be mapped to the maximal or additive algebras on the null-line for the union of the region and their reflection image through the origin. We studied the region arrangement on the half-line consisting of one interval attached to the origin and another interval extending to infinity, for which its dual algebra is additive. Our identification then implies that the known expression for the chiral current must coincide with the one for the scalar on the half-line. Besides being interesting on its own, this novel RMI coincides with the difference between the one corresponding to the massless scalar and (half of) the Maxwell fields for concentrical spherical regions. 

We studied our result in the limit of both short and long distance between the intervals. The former give universal information of the theory: the mutual information is expected to acquire a logarithmic divergent term proportional to the anomaly coefficient. We obtain the known mismatch for the Maxwell field that is due to the absence of charged fields. It is to be noted that the presence of heavy charges that would repair the anomaly coefficient for the Maxwell field do not alter the function $I_n^{(\text{M})}(\eta)$ of the free field up to very small values of $1-\eta\ll 1$ depending on the actual values of the charge masses and copling constants. 
Also, a subleading double logarithmic divergence appears at short distances, which is attributed to the IR divergence of the massless scalar field in $1+1$ dimensions.

In the opposite limit, we computed a series expansion in powers of the inverse of the separation distance. This type of expansion for the RMI of a CFT encodes information about the spectrum of the theory. This allowed us to crosscheck our result since the expansions for the scalar and Maxwell fields are known for the lowest orders (for the later only for $n=1$), and their difference precisely matches our expansion. Furthermore, we could use our result for generic $n$ in order to find the leading term in this expansion for the Maxwell field and $n\neq 1$, which had not been reported in the literature yet.

The long distance expansion is also interesting because we found out that the power series does not have a finite radius of convergence for every value of $n$. For integer $n> 1$ the series is convergent, which is expected since in this case it corresponds to the OPE series for localized operators in a CFT. For $n=1$, as well as for non-integer values, the expansion is only asymptotical. At a technical level this is allowed because for these values of $n$ the series can get contributions from an arbitrary number of copies. 

An interesting open question for the future is the Borel summability of this asymptotic expansion. This may indicate if there are non-perturbative contributions to the RMI of the model for these values of $n$. 
Another important question in this regard is about the physical origin of this behavior and how ubiquitous it is. In the $d=2$ case, we know that the RMI of the free fermion is convergent for any $n$, having a simple linear behavior in $n$. A prominent algebraic difference between the current and the fermion is that the former is not complete, i.e., it does not obey Haag duality for multi-interval regions. One may speculate that this might be the origin of the phenomena, and expect to encounter the same feature for other complete and non complete theories. However, in the $d=4$ case, both the Maxwell field and the scalar are Haag dual for multi-component spherical regions, though the Maxwell field is non complete for torus like regions. This last non completeness is in fact behind the mismatch of the anomaly coefficient with the logarithmic term. It would be a valuable information in this sense to know which function, the Maxwell or the scalar one, or both, is non convergent.     

It seems plausible that the difference of other RMI for free fields in higher dimensions can be solved by dimensional reduction as the present case. A promising case is the Rarita Schwinger field compared with a massless Dirac field.    

Finally, another direction in which our work might be extended is computing other information-theoretic quantities for the scalar on the half-line, such as the modular Hamiltonian or the modular conjugation. Moreover, since the model on the half-line is known to coincide with the zero angular momentum mode of a scalar field in $d=4$ dimensions, doing so could even help to better understand some aspects of entanglement for these theories in higher dimensions. 


\section*{Acknowledgements}

We thank Philip Argyres for fruitful discussions and encouragement.
This work is partially supported by CONICET, CNEA and Universidad Nacional  de Cuyo, Argentina.


\bibliography{EE}{}

\providecommand{\href}[2]{#2}\begingroup\raggedright\begin{thebibliography}{10}

\bibitem{Casini:2015woa}
H.~Casini, M.~Huerta, R.~Myers, and A.~Yale, ``{Mutual information and the
  F-theorem},'' \href{http://dx.doi.org/10.1007/JHEP10(2015)003}{{\em JHEP}
  {\bfseries 10} (2015) 003},
\href{http://arxiv.org/abs/1506.06195}{{\ttfamily arXiv:1506.06195 [hep-th]}}.

\bibitem{Calabrese:2010he}
P.~Calabrese, J.~Cardy, and E.~Tonni, ``{Entanglement entropy of two disjoint
  intervals in conformal field theory II},''
  \href{http://dx.doi.org/10.1088/1742-5468/2011/01/P01021}{{\em J. Stat.
  Mech.} {\bfseries 1101} (2011) P01021},
\href{http://arxiv.org/abs/1011.5482}{{\ttfamily arXiv:1011.5482 [hep-th]}}.

\bibitem{Cardy:2013nua}
J.~Cardy, ``{Some results on the mutual information of disjoint regions in
  higher dimensions},''
  \href{http://dx.doi.org/10.1088/1751-8113/46/28/285402}{{\em J. Phys. A}
  {\bfseries 46} (2013) 285402},
  \href{http://arxiv.org/abs/1304.7985}{{\ttfamily arXiv:1304.7985 [hep-th]}}.

\bibitem{Calabrese:2009ez}
P.~Calabrese, J.~Cardy, and E.~Tonni, ``{Entanglement entropy of two disjoint
  intervals in conformal field theory},''
  \href{http://dx.doi.org/10.1088/1742-5468/2009/11/P11001}{{\em J. Stat.
  Mech.} {\bfseries 0911} (2009) P11001},
\href{http://arxiv.org/abs/0905.2069}{{\ttfamily arXiv:0905.2069 [hep-th]}}.

\bibitem{Mintchev:2020uom}
M.~Mintchev and E.~Tonni, ``{Modular Hamiltonians for the massless Dirac field
  in the presence of a boundary},''
  \href{http://dx.doi.org/10.1007/JHEP03(2021)204}{{\em JHEP} {\bfseries 03}
  (2021) 204}, \href{http://arxiv.org/abs/2012.00703}{{\ttfamily
  arXiv:2012.00703 [hep-th]}}.

\bibitem{Estienne:2023ekf}
B.~Estienne, Y.~Ikhlef, A.~Rotaru, and E.~Tonni, ``{Entanglement entropies of
  an interval for the massless scalar field in the presence of a boundary},''
  \href{http://dx.doi.org/10.1007/JHEP05(2024)236}{{\em JHEP} {\bfseries 05}
  (2024) 236}, \href{http://arxiv.org/abs/2308.00614}{{\ttfamily
  arXiv:2308.00614 [hep-th]}}.

\bibitem{furukawa2009mutual}
S.~Furukawa, V.~Pasquier, and J.~Shiraishi, ``{Mutual Information and
  Compactification Radius in a c=1 Critical Phase in One Dimension},''
  \href{http://dx.doi.org/10.1103/PhysRevLett.102.170602}{{\em Phys. Rev.
  Lett.} {\bfseries 102} (2009) 170602},
  \href{http://arxiv.org/abs/0809.5113}{{\ttfamily arXiv:0809.5113
  [cond-mat.stat-mech]}}.

\bibitem{Headrick:2010zt}
M.~Headrick, ``{Entanglement Renyi entropies in holographic theories},''
  \href{http://dx.doi.org/10.1103/PhysRevD.82.126010}{{\em Phys. Rev.}
  {\bfseries D82} (2010) 126010},
\href{http://arxiv.org/abs/1006.0047}{{\ttfamily arXiv:1006.0047 [hep-th]}}.

\bibitem{casini2005entanglement}
H.~Casini, C.~D. Fosco, and M.~Huerta, ``{Entanglement and alpha entropies for
  a massive Dirac field in two dimensions},''
  \href{http://dx.doi.org/10.1088/1742-5468/2005/07/P07007}{{\em J. Stat.
  Mech.} {\bfseries 0507} (2005) P07007},
  \href{http://arxiv.org/abs/cond-mat/0505563}{{\ttfamily
  arXiv:cond-mat/0505563}}.

\bibitem{Arias:2018tmw}
R.~E. Arias, H.~Casini, M.~Huerta, and D.~Pontello, ``{Entropy and modular
  Hamiltonian for a free chiral scalar in two intervals},''
  \href{http://dx.doi.org/10.1103/PhysRevD.98.125008}{{\em Phys. Rev. D}
  {\bfseries 98} no.~12, (2018) 125008},
  \href{http://arxiv.org/abs/1809.00026}{{\ttfamily arXiv:1809.00026
  [hep-th]}}.

\bibitem{hirata2007ads}
T.~Hirata and T.~Takayanagi, ``{AdS/CFT and strong subadditivity of
  entanglement entropy},''
  \href{http://dx.doi.org/10.1088/1126-6708/2007/02/042}{{\em JHEP} {\bfseries
  02} (2007) 042}, \href{http://arxiv.org/abs/hep-th/0608213}{{\ttfamily
  arXiv:hep-th/0608213}}.

\bibitem{agon2016quantum}
C.~Ag{\'o}n and T.~Faulkner, ``{Quantum Corrections to Holographic Mutual
  Information},'' \href{http://dx.doi.org/10.1007/JHEP08(2016)118}{{\em JHEP}
  {\bfseries 08} (2016) 118}, \href{http://arxiv.org/abs/1511.07462}{{\ttfamily
  arXiv:1511.07462 [hep-th]}}.

\bibitem{agon2025mutual}
C.~A. Agon, H.~Casini, U.~G{\"u}rsoy, and G.~Planella~Planas, ``{Mutual
  information from modular flow in CFTs},''
  \href{http://dx.doi.org/10.1007/JHEP08(2025)176}{{\em JHEP} {\bfseries 08}
  (2025) 176}, \href{http://arxiv.org/abs/2409.01406}{{\ttfamily
  arXiv:2409.01406 [hep-th]}}.

\bibitem{Chen:2016mya}
B.~Chen and J.~Long, ``{R\'enyi mutual information for a free scalar field in
  even dimensions},'' \href{http://dx.doi.org/10.1103/PhysRevD.96.045006}{{\em
  Phys. Rev.} {\bfseries D96} no.~4, (2017) 045006},
\href{http://arxiv.org/abs/1612.00114}{{\ttfamily arXiv:1612.00114 [hep-th]}}.

\bibitem{chen2013short}
B.~Chen and J.-J. Zhang, ``{On short interval expansion of R{\'e}nyi
  entropy},'' \href{http://dx.doi.org/10.1007/JHEP11(2013)164}{{\em JHEP}
  {\bfseries 11} (2013) 164}, \href{http://arxiv.org/abs/1309.5453}{{\ttfamily
  arXiv:1309.5453 [hep-th]}}.

\bibitem{chen2017mutual}
B.~Chen, L.~Chen, P.-x. Hao, and J.~Long, ``{On the Mutual Information in
  Conformal Field Theory},''
  \href{http://dx.doi.org/10.1007/JHEP06(2017)096}{{\em JHEP} {\bfseries 06}
  (2017) 096}, \href{http://arxiv.org/abs/1704.03692}{{\ttfamily
  arXiv:1704.03692 [hep-th]}}.

\bibitem{Casini:2021raa}
H.~Casini, E.~Test{\'e}, and G.~Torroba, ``{Mutual information superadditivity
  and unitarity bounds},''
  \href{http://dx.doi.org/10.1007/JHEP09(2021)046}{{\em JHEP} {\bfseries 09}
  (2021) 046}, \href{http://arxiv.org/abs/2103.15847}{{\ttfamily
  arXiv:2103.15847 [hep-th]}}.

\bibitem{agon2022tripartite}
C.~A. Ag{\'o}n, P.~Bueno, and H.~Casini, ``{Tripartite information at long
  distances},'' \href{http://dx.doi.org/10.21468/SciPostPhys.12.5.153}{{\em
  SciPost Phys.} {\bfseries 12} no.~5, (2022) 153},
  \href{http://arxiv.org/abs/2109.09179}{{\ttfamily arXiv:2109.09179
  [hep-th]}}.

\bibitem{casini2020logarithmic}
H.~Casini, M.~Huerta, J.~M. Mag{\'a}n, and D.~Pontello, ``{Logarithmic
  coefficient of the entanglement entropy of a Maxwell field},''
  \href{http://dx.doi.org/10.1103/PhysRevD.101.065020}{{\em Phys. Rev. D}
  {\bfseries 101} no.~6, (2020) 065020},
  \href{http://arxiv.org/abs/1911.00529}{{\ttfamily arXiv:1911.00529
  [hep-th]}}.

\bibitem{Casini:2015dsg}
H.~Casini and M.~Huerta, ``{Entanglement entropy of a Maxwell field on the
  sphere},'' \href{http://dx.doi.org/10.1103/PhysRevD.93.105031}{{\em Phys.
  Rev. D} {\bfseries 93} no.~10, (2016) 105031},
  \href{http://arxiv.org/abs/1512.06182}{{\ttfamily arXiv:1512.06182
  [hep-th]}}.

\bibitem{Casini:2019kex}
H.~Casini, M.~Huerta, J.~M. Mag{\'a}n, and D.~Pontello, ``{Entanglement entropy
  and superselection sectors. Part I. Global symmetries},''
  \href{http://dx.doi.org/10.1007/JHEP02(2020)014}{{\em JHEP} {\bfseries 02}
  (2020) 014}, \href{http://arxiv.org/abs/1905.10487}{{\ttfamily
  arXiv:1905.10487 [hep-th]}}.

\bibitem{Casini:2021zgr}
H.~Casini and J.~M. Magan, ``{On completeness and generalized symmetries in
  quantum field theory},''
  \href{http://dx.doi.org/10.1142/S0217732321300251}{{\em Mod. Phys. Lett. A}
  {\bfseries 36} no.~36, (2021) 2130025},
  \href{http://arxiv.org/abs/2110.11358}{{\ttfamily arXiv:2110.11358
  [hep-th]}}.

\bibitem{Huerta:2022tpq}
M.~Huerta and G.~van~der Velde, ``{Modular Hamiltonian of the scalar in the
  semi infinite line: dimensional reduction for spherically symmetric
  regions},'' \href{http://dx.doi.org/10.1007/JHEP06(2023)097}{{\em JHEP}
  {\bfseries 06} (2023) 097}, \href{http://arxiv.org/abs/2301.00294}{{\ttfamily
  arXiv:2301.00294 [hep-th]}}.

\bibitem{Buchholz:1990ew}
D.~Buchholz and H.~Schulz-Mirbach, ``{Haag Duality in Conformal Quantum Field
  Theory},'' \href{http://dx.doi.org/10.1142/S0129055X90000053}{{\em Rev. Math.
  Phys.} {\bfseries 2} (1990) 105--125}.

\bibitem{Garbarz:2021rqu}
A.~Garbarz and G.~Palau, ``{A note on Haag duality},''
  \href{http://dx.doi.org/10.1016/j.nuclphysb.2022.115797}{{\em Nucl. Phys. B}
  {\bfseries 980} (2022) 115797},
  \href{http://arxiv.org/abs/2108.01257}{{\ttfamily arXiv:2108.01257
  [hep-th]}}.

\bibitem{ECKMANN19731}
J.-P. Eckmann and K.~Osterwalder, ``An application of tomita's theory of
  modular hilbert algebras: Duality for free bose fields,''
  \href{http://dx.doi.org/https://doi.org/10.1016/0022-1236(73)90062-1}{{\em
  Journal of Functional Analysis} {\bfseries 13} no.~1, (1973) 1--12}.
  \url{https://www.sciencedirect.com/science/article/pii/0022123673900621}.

\bibitem{Calabrese:2009qy}
P.~Calabrese and J.~Cardy, ``{Entanglement entropy and conformal field
  theory},'' \href{http://dx.doi.org/10.1088/1751-8113/42/50/504005}{{\em J.
  Phys.} {\bfseries A42} (2009) 504005},
\href{http://arxiv.org/abs/0905.4013}{{\ttfamily arXiv:0905.4013
  [cond-mat.stat-mech]}}.

\bibitem{Casini:2009sr}
H.~Casini and M.~Huerta, ``{Entanglement entropy in free quantum field
  theory},'' \href{http://dx.doi.org/10.1088/1751-8113/42/50/504007}{{\em J.
  Phys. A} {\bfseries 42} (2009) 504007},
  \href{http://arxiv.org/abs/0905.2562}{{\ttfamily arXiv:0905.2562 [hep-th]}}.

\bibitem{casini:2010kt}
H.~Casini and M.~Huerta, ``{Entanglement entropy for the n-sphere},''
  \href{http://dx.doi.org/10.1016/j.physletb.2010.09.054}{{\em Phys. Lett. B}
  {\bfseries 694} (2011) 167--171},
  \href{http://arxiv.org/abs/1007.1813}{{\ttfamily arXiv:1007.1813 [hep-th]}}.

\bibitem{Dowker:2010bu}
J.~S. Dowker, ``{Entanglement entropy for even spheres},''
  \href{http://arxiv.org/abs/1009.3854}{{\ttfamily arXiv:1009.3854 [hep-th]}}.

\bibitem{boyd_devils_1999}
J.~P. Boyd, ``The {Devil}'s {Invention}: {Asymptotic}, {Superasymptotic} and
  {Hyperasymptotic} {Series},''
  \href{http://dx.doi.org/10.1023/A:1006145903624}{{\em Acta Applicandae
  Mathematica} {\bfseries 56} no.~1, (Mar., 1999) 1--98}.
  \url{https://doi.org/10.1023/A:1006145903624}.

\bibitem{Cardy:2007mb}
J.~L. Cardy, O.~A. Castro-Alvaredo, and B.~Doyon, ``{Form factors of
  branch-point twist fields in quantum integrable models and entanglement
  entropy},'' \href{http://dx.doi.org/10.1007/s10955-007-9422-x}{{\em J.
  Statist. Phys.} {\bfseries 130} (2008) 129--168},
  \href{http://arxiv.org/abs/0706.3384}{{\ttfamily arXiv:0706.3384 [hep-th]}}.

\bibitem{long2016co}
J.~Long, ``{On co-dimension two defect operators},''
  \href{http://arxiv.org/abs/1611.02485}{{\ttfamily arXiv:1611.02485
  [hep-th]}}.

\end{thebibliography}\endgroup
\bibliographystyle{utphys}

\end{document}